# Design and Fabrication of Coaxial Dual Core Optical Fiber Fan-in Device

Yuhong Ma, Shitai Yang[#], Libo Yuan[✉]

With the rapid development of information and communication technologies in recent years, the transmission capacity of single-core optical fibers has nearly reached its physical limit. Space-division multiplexing based on multi-core fibers offers an effective solution to this bottleneck. Multi-core fibers feature high integration and large transmission capacity, and their unique structural characteristics also give them special value in fiber-optic sensing applications. Among various types of multi-core fibers, coaxial dual-core fibers (CDCFs) have shown promising performance in particle trapping, signal emission, and spectral analysis. To enable reliable interconnection between different types of multi-core fibers and single-core fiber arrays, this paper presents the design and fabrication of a fan-in device for coaxial dual-core fibers with different core diameters. The proposed method relies solely on cold-processing techniques and does not require any fusion splicing or thermal processing. The device is implemented on a V-groove substrate. Through structural design, fabrication, and experimental characterization, the average insertion loss of the ring core and the central core at a wavelength of 980 nm is measured to be 2.15 dB and 1.25 dB, respectively, demonstrating the successful fabrication of a coaxial dual-core fan-in device.

## 1 Introduction

Since Arthur Ashkin et al. first proposed the optical tweezer technique in 1986, it has evolved into a key tool in numerous interdisciplinary fields due to its advantage of non-contact, precise manipulation[1]. Over the past few decades, fiber-optic tweezer technology has also made significant progress. Among these developments, specialty fibers with an annular core (AC) have demonstrated excellent particle trapping performance[2]. In practical applications, the output end face of AC fibers is typically micro-machined into a tapered or spherical lens structure[3] to achieve efficient focusing of the optical field. The focused annular beam can form a stable three-dimensional optical potential well, enabling robust trapping of individual cells or micrometer-sized particles. Currently, such annular-core fiber tweezers have been widely employed in cutting-edge research, including single-cell trapping[4], piconewton level weak force sensing[5], microdroplet resonator manipulation[6], and microfluidic dye lasers[7].To further expand the degrees of manipulation and enhance system integration, a coaxial dual-core fiber (CDCF) composed of a nested annular core (AC) and central core (CC) has been developed. In this multiplexed structure, the outer AC retains its fundamental role of generating an optical potential well for particle trapping, while the additional CC introduces two new physical mechanisms for manipulation.First, the output beam transmitted through the CC can be used to precisely fine-tune the axial radiation force applied to the trapped particle. When the optical power in the center channel reaches a specific threshold, the trapped particle can even overcome the potential well confinement and be directionally ejected like a "bullet," providing a new approach for the dynamic transport of microparticles[8]. Second, the CC can serve as a high-efficiency backward signal collection channel, enabling in situ reception and transmission of Raman scattering or fluorescence signals from the particle[9]. This capability holds significant academic value for real-time spectral analysis and biochemical detection of single cells[10].

As with other types of specialty fibers, achieving low-loss interconnection of CDCFs and efficient coupling with standard optical communication components is a key technological bottleneck that limits their transition to practical applications[11]. For basic direct connections, the CDCF possesses a perfectly axisymmetric transverse geometry, allowing its end-face interconnection and link-loss evaluation to be fully compatible with existing commercial fusion splicers, thereby enabling low-loss splicing comparable to that of conventional single-mode fibers. However, in complex system configurations, the central challenge lies in how to realize independent excitation and demultiplexing of the optical power in the AC and CC, respectively.

In early explorations, researchers attempted to use the traditional fused biconical taper process to fabricate fiber couplers for optical circuit integration. However, this approach is limited by the physical mechanism of mode field coupling, making it difficult to achieve free and decoupled control over the optical power in the two independent cores during practical operation[12]. Moreover, the thermal fusion method can also cause structural changes in the CDCF, leading to errors and high losses in the fabricated devices[13]. To address this challenge, subsequent studies proposed a side-polishing-based optical coupling scheme: by precisely grinding the cladding sides of a single-mode fiber (SMF) and the CDCF and physically binding them together, the evanescent wave effect is utilized to directionally couple the optical field from the SMF into the annular core. Meanwhile, the CC establishes an optical channel by directly fusion-splicing its end face with another SMF[14]. Using this fan-in coupling device based on the side-polishing technique, the research team successfully applied the CDCF in optical experiments for trapping and directionally ejecting single polystyrene microspheres. Nevertheless, due to the microscopic roughness of the polished interface and imperfect mode matching, the energy coupling efficiency into the annular core via this side-polishing method is generally low. In addition, the physical binding of the mating surfaces is highly sensitive to environmental vibrations and temperature fluctuations, making it difficult to ensure the long-term stability of such fan-in devices. Therefore, it is imperative to explore more stable and efficient fan-in integration solutions to fully overcome the engineering limitations of coaxial dual-core fibers in multifunctional optical tweezer systems.

This paper provides a detailed introduction to the structural parameters of two types of CDCFs. The coupling efficiencies of different modes in the AC and CC are discussed. Subsequently, a CDCF fan-in device is proposed and fabricated, which consists of a fiber-fixing quartz substrate, a quartz cover plate, a CDCF, and two single-core fibers. By physically polishing one of the single-core fibers and applying plasma sputter coating, the light beam is injected into the AC; the other single-core fiber is physically butt-coupled to the CDCF to achieve beam injection into the CC. Factors causing insertion loss in the fan-in device, such as the polishing angle, are simulated and discussed. Finally, the fan-in device is fabricated, and its performance is evaluated by measuring the output optical field and insertion loss. With the gradual increase in the CC diameter of the CDCF, the method proposed in this paper avoids the need for thermal fusion processing of the fiber, while still enabling the fabrication of CDCF fan-in devices under various CC diameter conditions. Moreover, this method is also applicable to the fabrication of fan-in devices for other types of multi-core fibers, such as dual-core fibers and three-core fibers, exhibiting a certain degree of universal applicability.

## 2 Coaxial dual core optical fiber

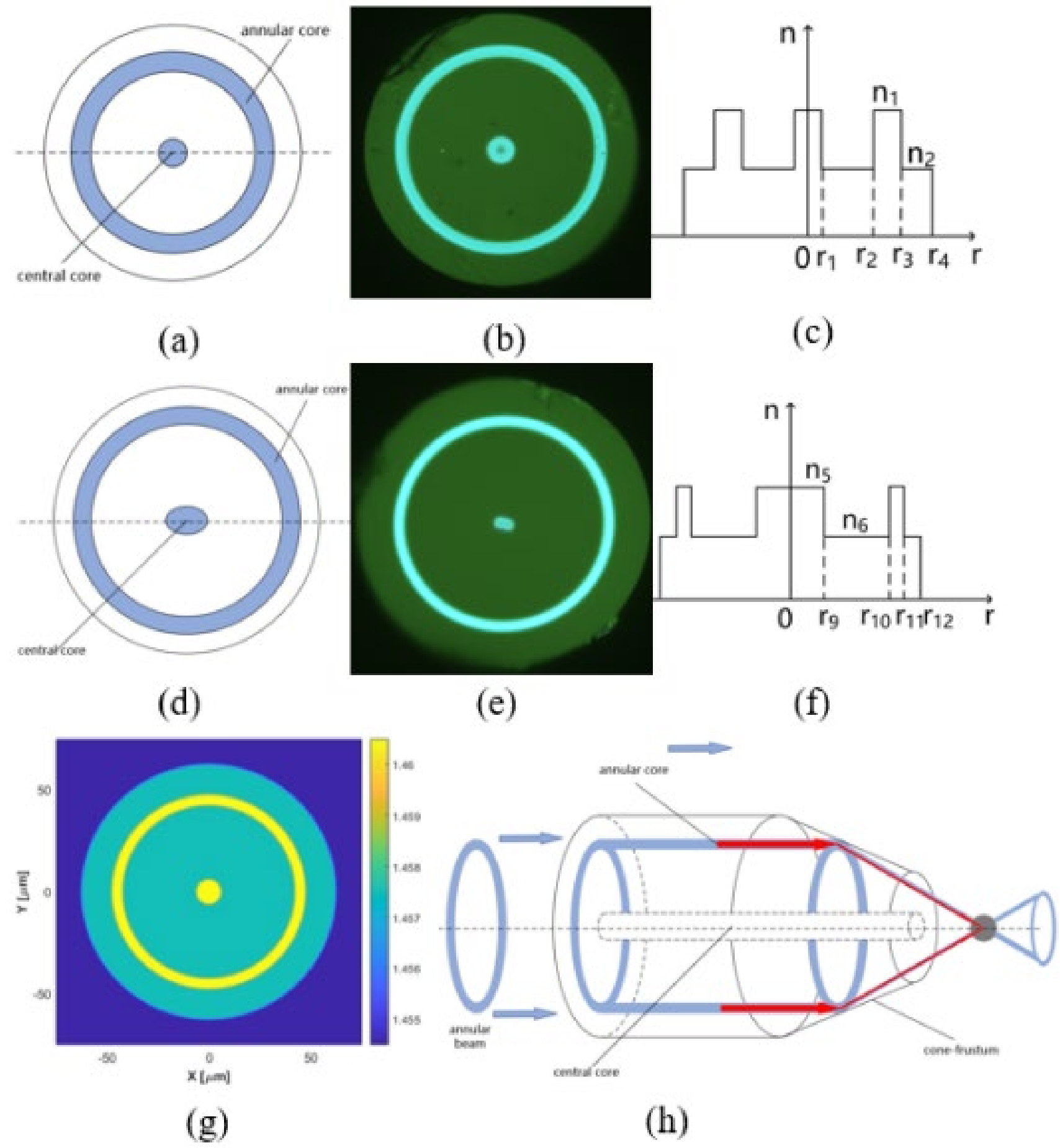

Fig. 1. (a) Cross-sectional schematic of a CDCF with a circular CC. (b) Microscope image of a CDCF with a circular CC. (c) Refractive-index (RI) profile of a CDCF with a circular CC. (d) Cross-sectional schematic of a CDCF with an elliptical CC. (e) Microscope image of a CDCF with an elliptical CC. (f) RI profile of a CDCF with an elliptical CC. (g) Measured RI profile of a CDCF with a circular CC using a refractive-index analyzer. (h) Schematic of particle trapping at the fiber end with a truncated-cone tip.

Figure 1(a) shows the structure of a CDCF. Figure 1(b) is a micrograph of the end face of this fiber. The geometric parameters of this CDCF are $r_1 = 6$ μm, $r_2 = 42.5$ μm, $r_3 = 47.5$ μm, and $r_4 = 62.5$ μm, respectively. The corresponding positional relationships and refractive index parameters are shown in Figure 1(c). Figure 1(c) displays the refractive index (RI) distribution along the dashed line in Figure 1(a), where the cladding RI is $n_2 = 1.4572$ (at 633 nm). The AC and CC of this fiber share the same RI value of $n_1 = 1.4602$. Figure 1(d) shows the structure of the second type of CDCF, and Figure 1(e) is a micrograph of its end face. Figure 1(f) presents the structural parameters and corresponding RI values along the dashed line in Figure 1(d), which are: $r_9 = 3$ μm, $r_{10} = 42.5$ μm, $r_{11} = 47.5$ μm, $r_{12} = 62.5$ μm, $n_5 = 1.4602$, and $n_6 = 1.4572$. The CC of this CDCF has an elliptical shape, with a major axis of 6 μm and a minor axis of 3 μm. The two-dimensional RI distribution of the CDCF shown in Figure 1(a) is presented in Figure 1(g), which was measured using a refractive index profiler (Photon Kinetics S14) and is consistent with the results shown in Figure 1(c). The two types of CDCFs have similar RI distributions, differing only in the size and structure of the CC.The primary application of CDCFs is in single-particle manipulation, particularly in single-cell trapping, manipulation, and related studies. Therefore, the tip of the CDCF is typically fabricated into a conical shape. Through this machined conical structure, the annular beam propagating in the AC is focused onto a single focal point, as illustrated in Figure 1(h). The conical tip is fabricated by grinding and polishing the fiber end, and its angle is designed based on simulation parameters to determine the optimal trapping force[15]. However, optical trapping is only one function of the CDCF probe. Because the CDCF possesses a CC structure, it also offers additional functionalities. For example, the CC can serve at least the following two functions.

1. In the optical tweezer system constructed using a CDCF, after a microscale target is stably trapped in

the potential well by the focused annular beam, the CC can serve as a power channel to introduce a Gaussian beam. By utilizing the axial radiation pressure, the particle can be controllably propelled away from the fiber tip. Through dynamic regulation of the input optical power in the center channel, precise characterization of the particle's propulsion speed and displacement distance can be achieved. This functional leap from "static trapping" to "dynamic transport" significantly enriches the manipulation dimension of fiber tweezers, offering substantial application value in fields such as microfluidic cell sorting, targeted drug delivery, and pharmacokinetic observation[16].

2. Furthermore, owing to the physical characteristics of fine diameter and high flexibility[17], the CDCF exhibits excellent potential for minimally invasive in vivo diagnosis and treatment, enabling in situ sensing deep within blood vessels or natural cavities[18]. In practical operation, the AC not only serves as the trapping channel for particle capture but also functions as the excitation light source pathway for Raman or fluorescence spectroscopy. Meanwhile, the center core is specifically dedicated to efficiently collecting the backscattered signals from the trapped target. This independent configuration of the physical channels achieves spatial separation between the excitation light field and the signal collection path, thereby effectively suppressing the crosstalk caused by endogenous fluorescence from the fiber matrix and significantly improving the signal-to-noise ratio of the spectral analysis[19]. This characteristic holds substantial engineering and practical value for achieving precise clinical diagnosis of diseases at the single-cell level.

## 3 Design of the dual core optical fiber fan-in device

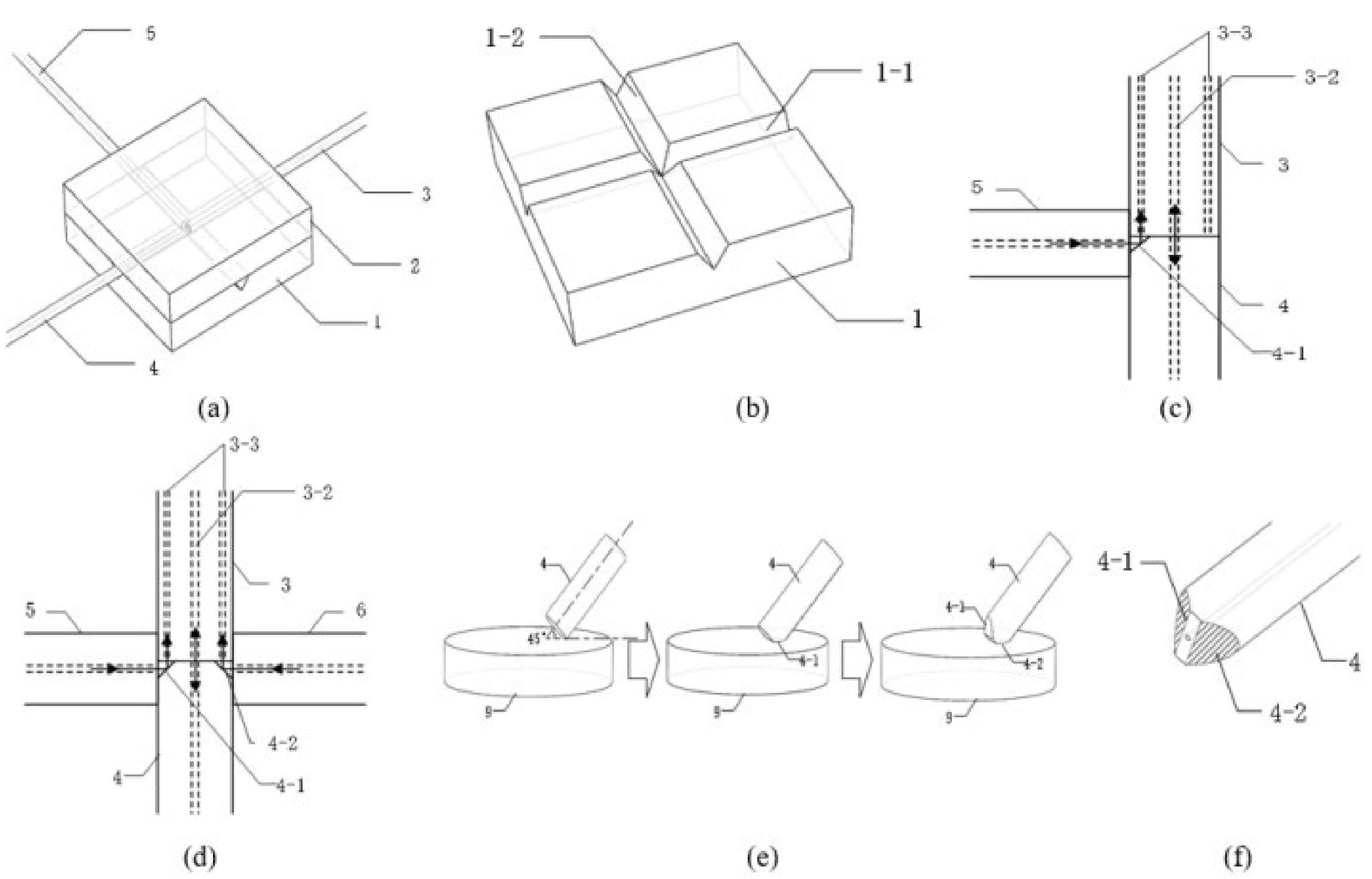

Fig. 2. (a) Schematic of the core structure of the CDCF. (b) Schematic of the quartz substrate structure. (c) Fiber positioning for injecting a single-spot Gaussian beam into the AC. (d) Fiber positioning for injecting a dual-spot Gaussian beam into the AC. (e) Illustration of the fiber polishing process. (f) Schematic of the fiber end-face after polishing.

In the second part of this paper, focusing on CDCF applications, we observe that CDCFs are primarily used for particle trapping, manipulation, and spectroscopic detection. Therefore, a fan-in device capable of independently routing light into both the AC and the CC is required. For the AC, the main objective is to achieve high injection efficiency and low loss so that the resulting trapping performance is optimized. For the CC, it is essential not only to ensure efficient forward transmission of light but also to maintain high efficiency during backward transmission for signal collection.To address the challenge of achieving an efficient and generalizable connection for different types of CDCFs, this paper presents a fabrication scheme for a fiber fan-in/fan-out (FIFO) device suitable for CDCFs with various CC sizes.

As shown in Fig. 2(a), the proposed CDCF FIFO device consists of a fiber-fixing quartz substrate 1, a quartz cover plate 2, a CDCF 3, a first single-core fiber 4, and a second single-core fiber 5. The CDCFs used here correspond to the three types described in Part II of this paper, all of which contain both a CC and an AC.The fiber-fixing quartz substrate 1 contains two mutually orthogonal V-grooves, as shown in Fig. 2(b). In the figure, 1-1 and 1-2 denote the two perpendicular V-groove structures. Figure 2(c) illustrates the fiber placement configuration corresponding to Fig. 2(a). The arrows indicate the direction of light propagation. The cleaved end of the CDCF is positioned in the first V-groove opposite the first single-core fiber, which has a 45° reflective facet. The cleaved end of the second single-core fiber is placed in the second V-groove.The CC of the CDCF is aligned with the core of the first single-core fiber to establish the central-core optical path. Light output from the second single-core fiber is reflected by the 45° facet of the first single-core fiber and then coupled into the AC of the CDCF, enabling independent injection into the annular core while maintaining a separate optical path for the CC. After the relative positions of the three fibers are precisely adjusted, the quartz cover plate 2 is applied to complete the encapsulation of the FIFO device.As shown in Fig. 2(d), a third single-core fiber 6 may also be placed symmetrically on the opposite side of the V-groove that holds the second single-core fiber 5. This third fiber is identical in structure to the second one. Light from both fibers can be reflected by the 45° facet of the first single-core fiber and injected into the AC of the CDCF. This configuration does not affect the alignment between the CC of the CDCF and the core of the first single-core fiber. The arrows in the figure indicate the direction of light propagation.Figure 2(e) illustrates the machining process for the first single-core fiber 4. The fiber is placed on a polishing plate 9 and rotated to a 45° angle. The polishing plate is then activated to machine the first 45° facet 4-1. Afterward, the fiber is rotated by 180° and processed again in the same manner to form the second 45° facet 4-2. The polishing plate is then replaced with a fine-polishing plate to smooth both 45° facets and improve their surface quality.Figure 2(f) shows the fully processed first single-core fiber. Both 45° facets are coated with a high-reflectivity film, enabling efficient coupling of light from the second and third single-core fibers into the AC of the CDCF.

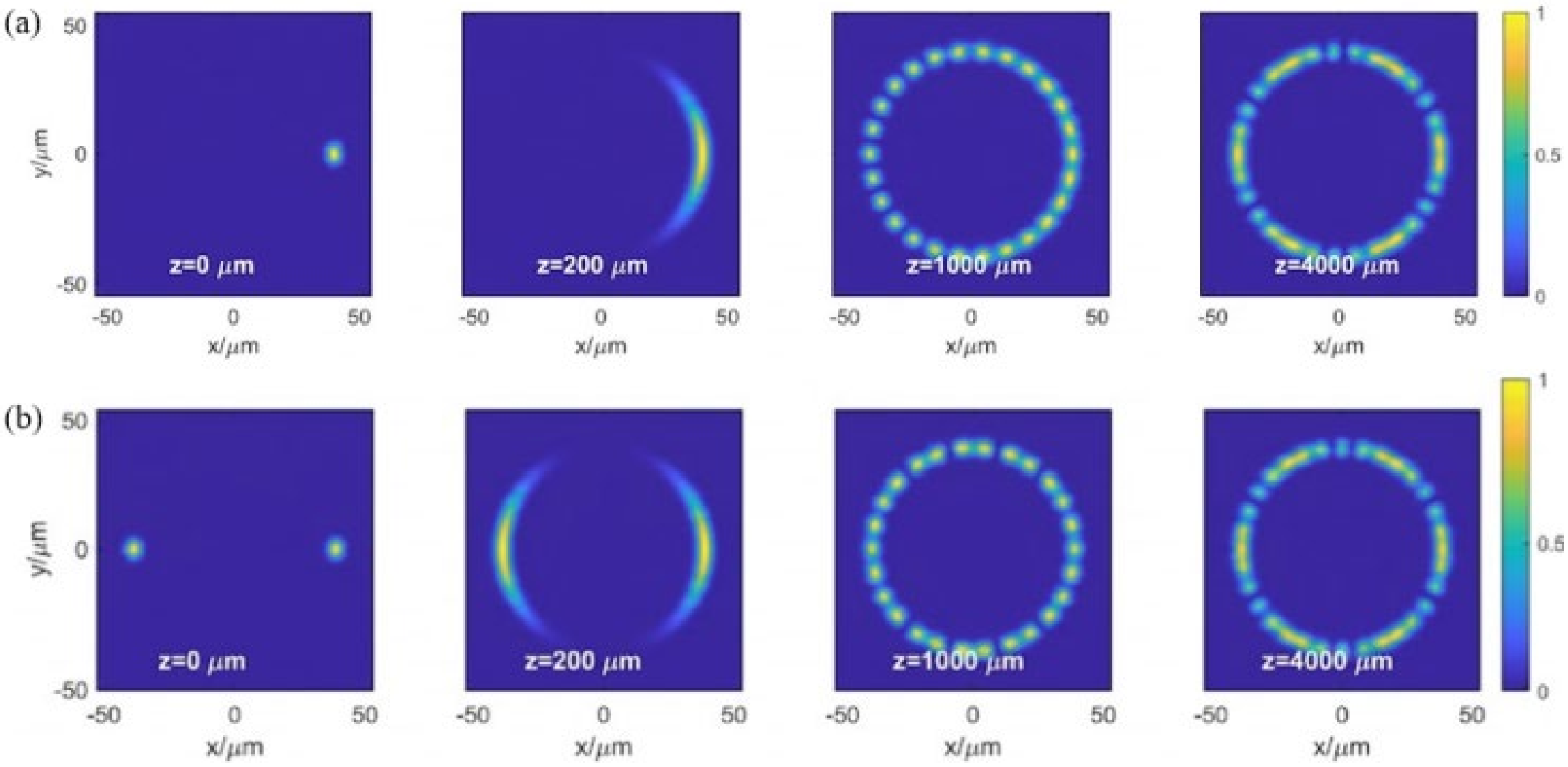

Fig. 3. (a) Evolution of the optical field in the AC with propagation distance for a single Gaussian beam injected into the AC. (b) Evolution of the optical field in the AC with propagation distance for two Gaussian beams injected into the AC.

As shown in Fig. 3, the evolution of the optical field in the AC with propagation distance can be obtained by numerical simulation based on the structural and refractive-index parameters of the annular core. Since the proposed device allows light to be coupled into the AC from one side or simultaneously from both sides, the outputs of the single-core fibers can be modeled as one or two Gaussian beams, corresponding to one or two point-source inputs to the AC. We use the beam-propagation method (BPM) to simulate the propagation of these beams in the AC.First, we consider the case where only one single-core fiber injects light into the AC, i.e., a single Gaussian beam input. The wavelength and mode-field diameter of this beam are 980 nm and 9 μm, respectively. Figure 3(a) shows the optical-field distribution in the AC at propagation distances z = 0 μm, 200 μm, 1000 μm, and 4000 μm for this single-beam excitation.As observed, the Gaussian field gradually spreads along the circular path of the AC. Once the AC becomes fully illuminated, interference occurs, leading to the formation of various evolving intensity patterns. When two single-core fibers inject light simultaneously—i.e., two Gaussian beam inputs—the field evolution during propagation is shown in Fig. 3(b). These results demonstrate that an appropriately chosen fundamental Gaussian mode can be efficiently coupled into the AC. Although the interference pattern continues to vary during propagation, the optical power remains uniformly distributed within the annular region, which does not affect the functional performance of the CDCF. In addition, when both single-core fibers inject light into the AC, different wavelengths may be launched from each fiber, enabling wavelength-division multiplexing.

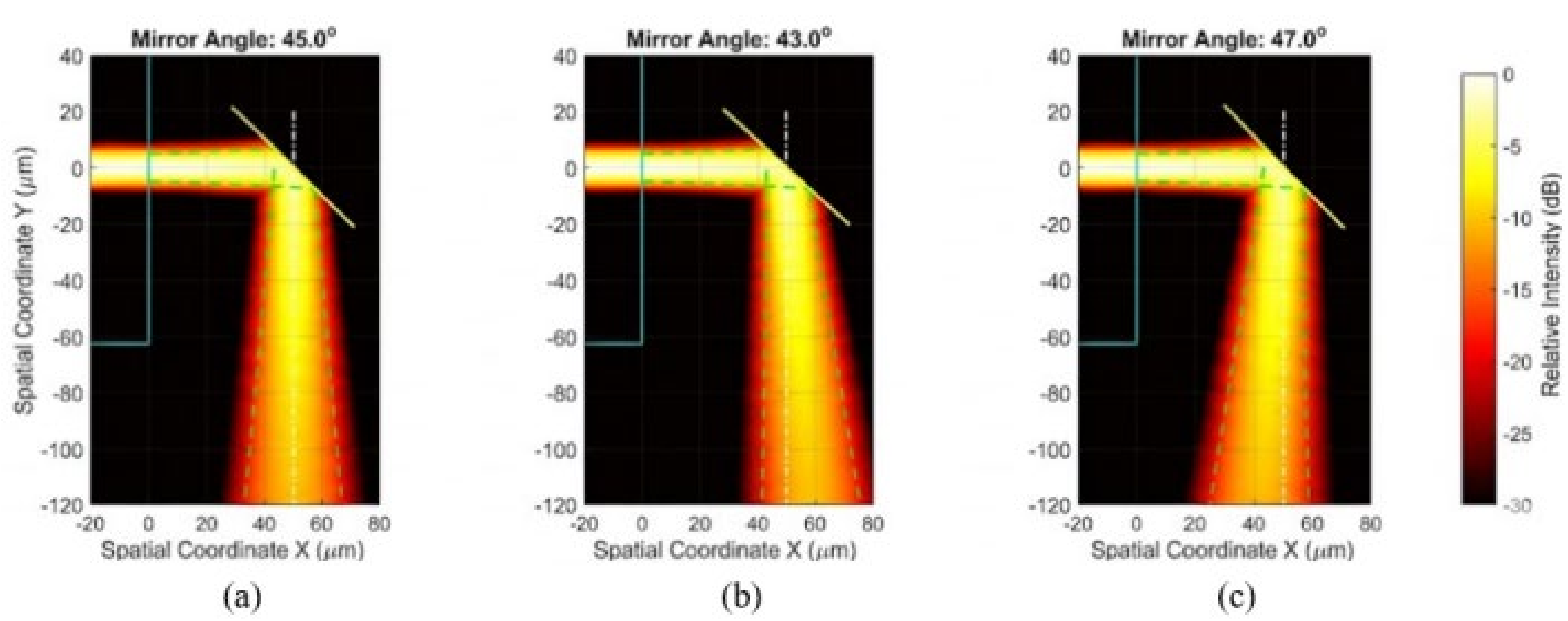

Fig. 4. (a) Beam propagation path and relative intensity with a reflective-facet angle of 45°. (b) Beam propagation path and relative intensity with a reflective-facet angle of 43°. (c) Beam propagation path and relative intensity with a reflective-facet angle of 47°.

Based on the proposed device design, we numerically analyze the dominant error source—namely, deviations in the polishing angle of the reflective facet. When the polishing angle of the first single-core fiber deviates from the ideal 45°, the coupling efficiency into the AC decreases significantly.We simulate the propagation path and relative intensity of a 980 nm beam reflected by facets with different angles. As shown in Fig. 4, for facet angles of 45°, 43°, and 47°, the reflected beam exhibits a pronounced trajectory shift when the angle deviates from 45°. This indicates that, after exiting the single-core fiber and being reflected by a misaligned mirror, the beam arrives at the cleaved end-face of the CDCF with a lateral offset. Owing to the mismatch between the beam position and the AC region—as well as the effect of differing mode-field diameters—the light spot can no longer be fully confined within the annular core, leading to a marked reduction in coupling efficiency.The core diameter and refractive index of the output end-faces of the second and third single-core fibers determine the mode-field diameter of their fundamental modes. After reflection by the 45° facet, the mode-field diameter of the reflected beam can no longer be stably controlled. As a result, the output beam spot cannot be accurately injected into the AC in the same way as in a conventional optical path. This prevents the formation of the desired output mode described earlier and leads to a substantial reduction in coupling efficiency.

## 4 Fabrication of the dual core optical fiber fan-in device

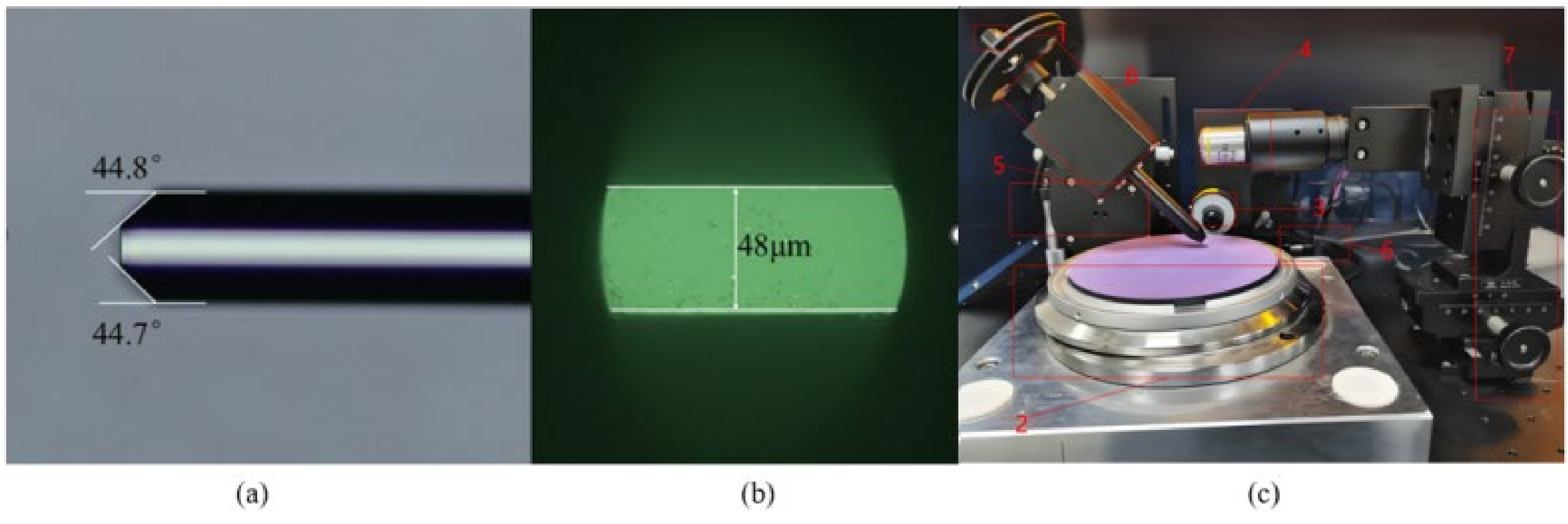

Fig. 5. (a) Microscope image of the first single-core fiber. (b) Corresponding end-face microscope image. (c) Photograph of the fiber-polishing machine.

Based on the structural requirements described in this work, the first single-core fiber must be machined into a symmetric 45° truncated-pyramid geometry to satisfy the fabrication specifications of the device. Figure 5(a) shows a side-view microscope image of such a symmetric 45° structure, and Fig. 5(b) shows the corresponding end-face image of the same fiber.Figure 5(c) presents a photograph of the polishing machine used for fiber processing. Component descriptions are as follows:1 is the fiber clamping device, which holds the fiber by threading it through the interior to the polishing wheel position and then clamping it in place.2 is the polishing wheel, whose rotation speed can be controlled to adjust the polishing rate of the fiber end face.3 is the side-view lighting and camera system used to observe and measure the polishing angle of the fiber.4 is the front-view lighting and camera system used to inspect the condition of the fiber end face.5 is an adjustable translation stage, which can be used to adjust the polishing angle and depth of the fiber, while the angle change can be monitored via the camera system in 3.6 is the adjustment translation stage for the side-view camera system 3, used to position the side camera at the appropriate location.7 is the adjustment translation stage for the front-view camera system 4, used to position the front camera at the appropriate location.8 is a coaxial rotation device that allows the fiber to pass through its center while maintaining coaxial alignment.When fabricating the symmetrical 45° prism-shaped structure shown in Figure 5(a), the fiber is first threaded through component 1 and then passed through component 8 to emerge from the other side. The front end of the fiber is located using the side-view camera system 3. By adjusting the translation stage 5, the fiber angle is set to 45°. Then, the polishing wheel 2 is activated to rotate, creating a 45° bevel on one side. Subsequently, the coaxial rotation device 8 is adjusted to rotate the fiber by 180° while maintaining coaxial alignment. The polishing wheel 2 is then activated again for the same duration to produce a 45° bevel that is perfectly symmetrical to the first one. Finally, through a polishing process, a prism-shaped fiber end with two symmetrical 45° bevels is obtained. After obtaining the symmetrical 45° prism structure as shown in the figure, this paper adopts a metal film deposition method to enhance the light reflection performance of the two 45° bevels on the fiber. Once the bevel coating step is completed, the first single-core fiber required for device fabrication is obtained[20].

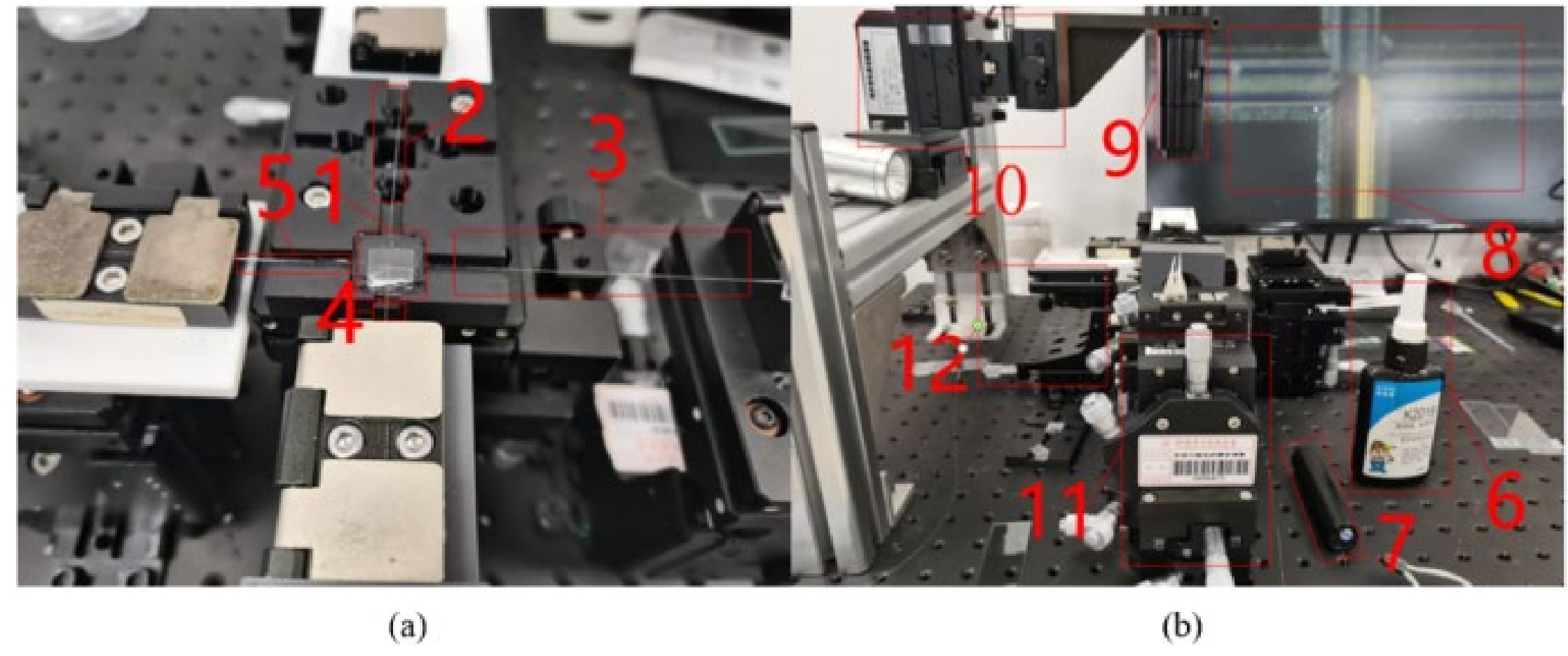

(a) (b)

Fig. 6. (a) Assembly setup for the fan-in device. (b) Alignment and testing setup for the fan-in device.

From the procedure described above, we can fabricate the required fiber-end 45° reflectors. Next, two single-mode fibers operating at 980 nm are cleaved flat to serve as the second and third single-core fibers. After cleaving the end face of the selected CDCF, the assembly and alignment of the fan-in device can begin. Figures 6(a) and 6(b) show the setup used for assembling and tuning the device.In the figure:1 is the V-groove glass substrate with a cover glass, after encapsulation.8 is the magnified view of 1 as seen through the camera system 9.2 is the CDCF.3 and 5 are the second and third single-core fibers, respectively.4 is the first single-core fiber, which has been processed to have a symmetrical 45° reflective end face.6 is the UV-curable adhesive used for encapsulation.7 is the UV curing lamp used for encapsulation.9 is the lens tube of the camera system.10 is the adjustment mount used to connect and position the camera system.11 is a six-axis translation stage used to fix and adjust the first single-core fiber. This stage allows the position and angle of the first single-core fiber to be adjusted to maximize the coupling efficiency into the annular core.12 is a five-axis translation stage used to fix the second single-core fiber. The same type of setup is also used to fix the third single-core fiber and the CDCF, and it can adjust the forward/backward position of the fibers.Beneath the entire assembly of 1 is a five-axis translation stage used to support the entire device. Figure 6(a) shows the fully assembled device after alignment and encapsulation. After adjusting each fiber to its appropriate position, the positions of the fibers in the V-groove are fixed using the UV-curable adhesive and UV curing lamp described above. The overall size of the device is within 15 mm.

Table.1. Insertion loss measurement results of the fan-in device fabricated from CDCF with circular CC.

| | | Fiber 1 (central core) | Fiber 2 (annular core) | Fiber 3 (annular core) |
|---|---|---|---|---|
| air | Insert loss | 1.442dB | 2.321dB | 2.218dB |
| RI matching liquid | Insert loss | 1.296dB | 2.204dB | 2.132dB |

Table.2. Insertion loss measurement results of the fan-in device fabricated from CDCF with elliptical CC.

| | | Fiber 1 (central core) | Fiber 2 (annular core) | Fiber 2 (annular core) |
|---|---|---|---|---|
| air | Insert loss | 1.343dB | 2.479dB | 2.358dB |
| RI matching liquid | Insert loss | 1.225dB | 2.338dB | 2.231dB |

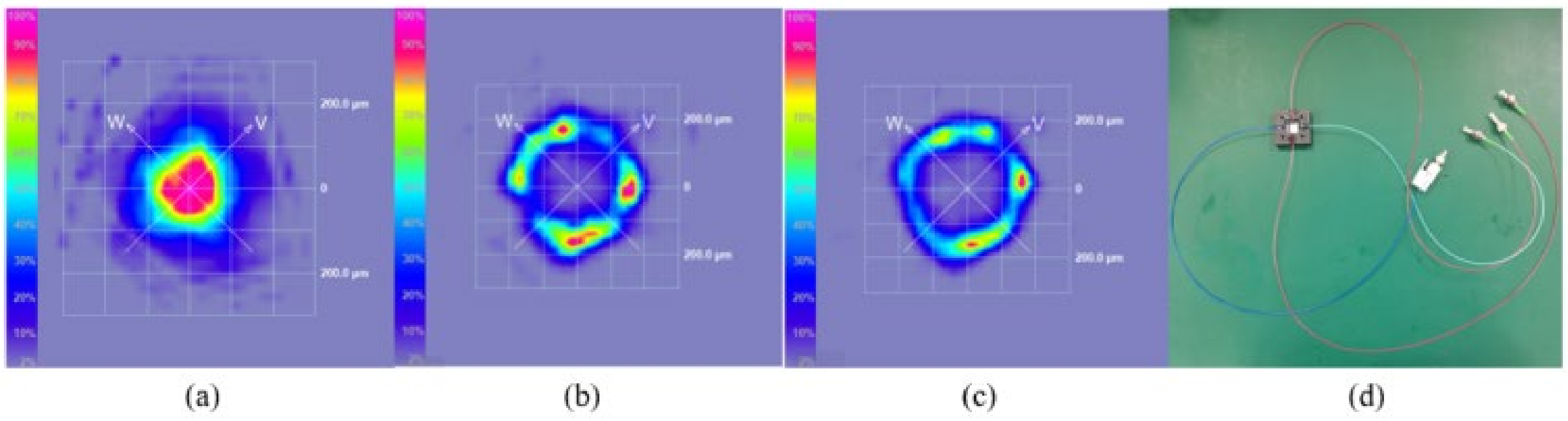

(a) (b) (c) (d)

Fig. 7. (a) Measured output field when light is injected into the CC. (b) Measured AC output field when light is injected through the second single-core fiber. (c) Measured AC output field when light is injected through the third single-core fiber. (d) Photograph of the fan-in device after encapsulation.

Performance characterization was conducted on the encapsulated devices, with the resulting data summarized in Tables 1 and 2. For the various CDCF configurations, the average insertion loss (IL) of the AC was measured at 2.15 dB. Although the two CDCFs differ in their overall structures, their CC components exhibit significant similarities, leading to a consistent IL performance with a mean value of 1.25 dB. using the CDCF with a circular CC as an example to illustrate the output field patterns. A 980 nm ASE source was used as the input for each SMF. By connecting the end of the CDCF to a beam analyzer (BA7-IR5, Duma Optronics Ltd.), the three output spots shown in Fig. 7 were obtained.Figure 7(a) shows the output spot pattern measured when light is injected into the CC and directly delivered to the beam analyzer. Figures 7(b) and 7(c) show the AC output patterns when light from the single-core fibers is coupled into the AC and then measured by the analyzer. These results match the expected field distribution of the CDCF.Loss measurements were also performed for different CDCF structures under two conditions: light propagation through air and through a refractive-index-matching liquid. As shown in the table, the use of the index-matching liquid significantly reduces insertion loss.Figure 7(d) shows the packaged fan-in device assembled using a 3D-printed plastic frame and detachable connectors, enabling stable operation in subsequent applications.

## 5 Conclusion

This work presents a fan-in device for CDCFs that achieves core-independent coupling without any thermal-fusion processing. Using CDCFs with two different CC structures as examples, we performed detailed measurements and analysis of their structural parameters and proposed the corresponding fan-in device design. Numerical simulations were conducted to clarify the light-propagation principles in the AC and to identify the fabrication-related factors responsible for increased insertion loss. The device was then fabricated, packaged, and experimentally tested.At a wavelength of 980 nm, the average insertion loss of the AC is below 2.48 dB, and that of the CC is below 1.45 dB. The elimination of thermal-fusion processing represents a key advantage of this new structure, ensuring that the CDCF geometry remains unaffected by thermal deformation. This approach provides a robust and versatile connection method for CDCFs with different CC designs and supports future multifunctional applications.